A novel method based on cross correlation maximization, for pattern matching by means of a single parameter. Application to the human voice.

Felipe Quiero+, Fabian Quintana and Leonardo Bennun*[c]

+ *Andres Bello University, Chile*

‡ *Concepcion University, Concepcion, Chile*

*Applied Physics Laboratory, Physics Department, Concepcion University, Concepcion, Chile.*
[c] *corresponding author:* [lbennun@udec.cl](mailto:lbennun@udec.cl)

Abstract: This work develops a cross correlation maximization technique, based on statistical concepts, for pattern matching purposes in time series. The technique analytically quantifies the extent of similitude between a known signal within a group of data, by means of a single parameter. Specifically, the method was applied to voice recognition problem, by selecting samples from a given individual's recordings of the 5 vowels, in Spanish. The frequency of acquisition of the data was 11.250 Hz. A certain distinctive interval was established from each vowel time series as a representative test function and it was compared both to itself and to the rest of the vowels by means of an algorithm, for a subsequent graphic illustration of the results.

We conclude that for a minimum distinctive length, the method meets resemblance between every vowel with itself, and also an irrefutable difference with the rest of the vowels for an estimate length of 30 points (~2 $10^{-3}$ s).

## 1. Introduction

There are standard procedures commonly used for the analysis of time series, which are widely applied for signal processing in such fields as communication, electronics and control sciences. These formulations have very diverse application areas, as many quantities in nature fluctuate in time. Examples are the weather, seismic waves, sunspots, heartbeat, plants and animals populations, stock market, etc.

Time series methods take into account possible internal structure in the data. They may be divided into two classes: time-domain methods and frequency-domain methods. The former include mathematical tools mainly based on autocorrelations, cross-correlations [1] and convolutions [2,3]; the latter include spectral analysis and recently wavelet analysis [4]. Both sorts of formulations can be directly related by the Fourier Transform [5].

A great deal of statistic procedures considers that the acquired data follow a certain type of mathematical model that is defined by an equation, in which some of its parameters are unknown. These are calculated or estimated from the information gathered in a well-designed analysis for such purpose. There are several procedures to estimate the model coefficients, or to estimate parameters of a distribution function assumed by the time of working with data series [6]. In order to evaluate the goodness of fit of a parameter within a statistic model, one need to know which is the distribution function where the sample data come from, or which is the most suitable distribution function for them. For that purpose, there are statistic validation techniques which enable to test the goodness of fit. One is the "Chi-square" test proposed by Pearson [7], highlighted from its beginning, used to assess adjustment of a categorical variable that takes discrete values, and another is the "Kolmogorov-Smirnov" test, more suitable for situations where the interest is placed one valuating the adjustment of variables taking continuous values [8]. Both methods can help checking goodness of fit and are based on showing the degree of relationship between the distribution obtained from sampling and the proposed theoretical distribution that the sample must supposedly follow. In spite of the huge success of these two methods, there are other tests to measure the goodness of fit. These are variations of the above-mentioned methods, and new choices.[9,10]

The main interest of this work is to implement a statistical method for characterization and recognition of human voice signals.

The voice signal is one of the most important tools for communication between human beings and their environment. Systems accurateness for voice recognition are affected by several parameters, such as: identification of discrete and continuous words, dependence on sender source, size of the word, used language, type of speech (colloquial), and environmental conditions. Generally, there are three usual methods in speech recognition: Dynamic Time Warping (DTW), Hidden Markov Model (HMM) and Artificial Neural Networks (ANNs) [11].

Dynamic time warping (DTW) is a technique that finds the optimal alignment between two time series if one time series may be warped non-linearly by stretching or shrinking it along its time axis. This warping between two time series can then be used to find corresponding regions between the two time series or to determine the similarity between them.[12]

Hidden Markov Models are finite automates, having a given number of states; passing from one state to another is made instantaneously at equally spaced time moments. At every pass from one state to another, the system generates observations, two processes are taking place: the transparent one, represented by the observations string (feature sequence), and the hidden one, which cannot be observed, represented by the state string [13,14].

Nowadays, ANNs are utilized [15] in wide ranges for their parallel distributed processing, distributed memories, error stability, and pattern learning distinguishing ability.[16] The Complexity of all these the systems increased when their generality rises. The biggest restriction of two first methods is their low speed for searching and comparing in models. But ANNs are faster, because output is resulted from multiplication of adjusted weights in present input. At present TDNN (Time-Delay Neural Network) is widely used in speech recognition.[17]

We have developed a methodology based on cross correlation maximization and statistical concepts, for time series analysis. This formulation opens new possibilities in the analysis of this kind of data, which were traditionally analyzed by time and frequency domain methods. The method was applied to the male voice recognition from Spanish vowels recordings of a male individual at a frequency of 11.250 Hz.

The mathematical model applied assumes that the data is affected by Gaussian statistical fluctuations. The algorithm developed is able to quantify the degree of similitude between two signals through one single parameter. A certain feature interval from each vowel data as test function was selected and later compared to itself and to the rest of the vowels. Neither, the data and the background were statistically checked in order to verify that they were affected by Gaussian fluctuations. Once the characteristic test function was established for each vowel, and in spite of not having carried out any statistical validation test, the results show that the proposed method finds definitive similitude between each vowel with itself and it is clearly differentiate from the rest of the vowels. Surprisingly, we conclude that for a minimum distinctive length, the method meets resemblance between every vowel with itself, and also a difference with the rest of the vowels for an estimate length of 30 points (~2 $10^{-3}$ s).

## 2. Time series analysis. Developed method.

In control, electronics, communications, etc. sciences, the most common data analysis are based on convolutions, cross-correlations, autocorrelation procedures, etc. A common procedure in signal processing, the cross-correlation, is a measure of the similarity of two waveforms as a function of a time-lag ($\tau$) applied to one of them [18,19]. This is also known as a *sliding dot product* or *inner-product*. It is commonly used to search a long duration signal for a shorter, known feature. It also has applications in pattern recognition.

For continuous functions, *f* and *g*, the cross-correlation (*cc*) is defined as:

$$cc(\tau) \equiv (f \otimes g)(\tau) \equiv \int_{-\infty}^{\infty} f^*(t) g(t+\tau) dt \qquad (1)$$

Where *f\** denotes the complex conjugate of *f*. The symbol $\otimes$ denotes the cross-correlation.

Similarly, for discrete functions, the cross-correlation is defined as**:**

$$cc(\tau) \equiv (f \otimes g)(\tau) \equiv \sum_{-\infty}^{\infty} f^*(t_j) g(t_j + \tau) \qquad (2)$$

By applying this techniques, an expected function *f(t)* and its respective intensity $\alpha$ may be determined in a given sequence of data *m(t)*, where *t* is time and *m* is usually expressed in Volts. The *cc* operation for this case is:

$$cc(\tau) \equiv (f \otimes m)(\tau) \equiv \sum_{-\infty}^{\infty} f^*(t_j) m(t_j + \tau) \qquad (3)$$

If the function *f(t)* and *m(t)*, are not related at all, the *cc(τ)* is zero.

Typical data *m(t)* is shown in Fig. 1, where a sound (the vowel "A" in Spanish) was recorded in the time domain.

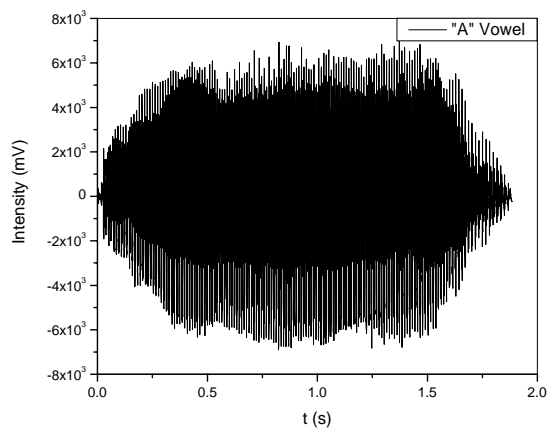

Figure 1. Spectrum obtained as a function of time from a recording of the vowel "A" in Spanish at 11.000 Hz.

In a common case, a discrete function $f(t_j)$ should be determined in a sequence of data $m(t_j)$ affected by an intensity *α*, as is depicted in Fig. 2.

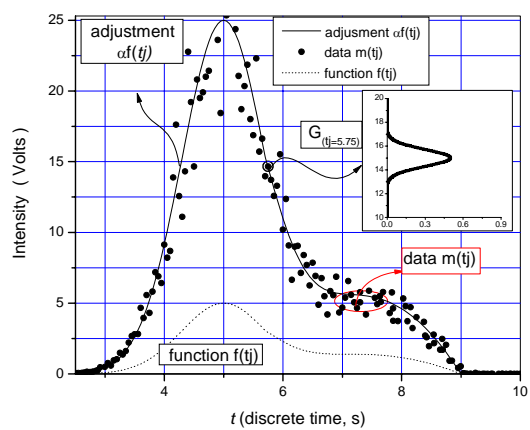

Figure 2. A discrete experimental data set, depending on time, is shown made of points $(t_j, m_j)$. In this set of data $m_j$, a function $f(t_j)$ must be found, affected by a parameter α. Each point $m_j$ is described by a random Gaussian function with centroid in the value $αf(t_j)$. A particular Gaussian function, at *t*=5.75 s, *G(j=5.75)*, is represented inside of the main graphic.

The sought function *f(t)* or its discrete version $f(t_j)$ is an arbitrary function. We can propose any kind of shape, properties or characteristics for it.

In the following we propose that each $m(t_j)$ (or simply $m_j$) datum is described by a random Gaussian function, with centroid at $αf(t_j)$ and variance $(σ_{bj})^2$, where the suffix *b* is due to the background or inherent statistical fluctuation affecting the data.

Each Gaussian function, $G(t_j)$ is described as:

$$G(\alpha f(t_j + \tau), \sigma_{bj}^2, m_j) = \frac{1}{\sqrt{2\pi}\sigma_{bj}} e^{-\frac{(m_j - \alpha f(t_j + \tau))^2}{2\sigma_{bj}^2}} \qquad (4)$$

The cross correlation between function *f(j)* and *m(j)* is expressed as:

$$cc(\tau) \equiv \sum_{-\infty}^{\infty} f(t_j) G(\alpha f(t_j + \tau), \sigma_{bj}^2, m_j) = \sum_{-\infty}^{\infty} f(t_j) \frac{1}{\sqrt{2\pi}\sigma_{bj}} e^{-\frac{(m_j - \alpha f(t_j + \tau))^2}{2\sigma_{bj}^2}} \qquad (5)$$

Here we are looking to maximize the cross-correlation with respect to $\alpha$. This is done by derivation of Eq. 5 with respect to $\alpha$, and equalizing to zero.

$$\frac{\partial}{\partial \alpha}(cc(\tau)) = \frac{\partial}{\partial \alpha} \sum_{-\infty}^{\infty} f(t_j) \frac{1}{\sqrt{2\pi}\sigma_{bj}} e^{-\frac{(m_j - \alpha f(t_j + \tau))^2}{2\sigma_{bj}^2}} = 0 \qquad (6)$$

We can insert the derivative operator inside the sum,

$$\frac{\partial}{\partial \alpha}(cc(\tau)) = \sum_{-\infty}^{\infty} \frac{\partial}{\partial \alpha}\left(f(t_j) \frac{1}{\sqrt{2\pi}\sigma_{bj}} e^{-\frac{(m_{ij} - \alpha f(t_j + \tau))^2}{2\sigma_{bj}^2}}\right) = 0 \qquad (7)$$

From the monotonic character of the function $ln(x)$, which always increases with $x$, it is equivalent to maximize $f(x)$ or $ln(f(x))$. Applying this property to Eq. 7, we obtain:

$$\frac{\partial}{\partial \alpha}(cc(\tau)) = \sum_{-\infty}^{\infty} \frac{\partial}{\partial \alpha}\left(\ln\left(\frac{f(t_j)}{\sqrt{2\pi}\sigma_{bj}}\right) - \frac{(m(t_j) - \alpha f(t_j + \tau))^2}{2\sigma_{bj}^2}\right) = 0 \qquad (8)$$

We notice that the first term in brackets does not depend on $\alpha$, so its derivative is zero. Then, Eq. 8 can be written as:

$$\frac{\partial(cc(\tau))}{\partial \alpha} = \sum_{-\infty}^{\infty} \frac{\partial}{\partial \alpha}\left(\frac{(m(t_j) - \alpha f(t_j + \tau))^2}{2\sigma_{bj}^2}\right) = 0 \qquad (9)$$

Finally, the $\alpha$ value which maximizes the $cc(\tau)$ is:

$$\alpha(\tau) = \frac{\sum_{j=1}^{n} \frac{m(t_j) f(t_j + \tau)}{\sigma_{bj}^2}}{\sum_{j=1}^{n} \frac{f(t_j + \tau)^2}{\sigma_{bj}^2}} \qquad (10)$$

By applying the variance operator in both terms of the Eq. 13, we obtain:

$$\Delta\alpha(\tau)^2 = \frac{1}{\sum_{j=1}^{n} \frac{f(t_j + \tau)^2}{\sigma_{bj}^2}} \qquad (11)$$

In Eqs. 10 and 11 the sums were reduced from ($-\infty$ to $+\infty$) to $j=1$ to $n$, since the usual number of data points is finite.

Finally, we obtain the following confidence interval for the parameter $\alpha$:

$$\alpha(\tau) = \frac{\sum_{j=1}^{n} \frac{m(t_j) f(t_j + \tau)}{\sigma_{bj}^2}}{\sum_{j=1}^{n} \frac{f(t_j + \tau)^2}{\sigma_{bj}^2}} \pm \frac{1}{\sqrt{\sum_{j=1}^{n} \frac{f(t_j + \tau)^2}{\sigma_{bj}^2}}} \qquad (12)$$

The $\sigma^2_{bj}$ value of the Eq. 4 (uncertainty or standard deviation of $m(t_j)$ in the $j^{th}$ channel), considers the inherent random variations of the data $m(t_j)$. In all of the formulation developed, we proposed that if we measure the $j^{th}$ channel many times (repeating over and over the same experiment, in the same controlled conditions), the Gaussian distribution obtained is depicted in Fig. 3. In estimating the overall uncertainty at each channel, it may be necessary to take each source of uncertainty and consider it separately. In practice, if the measurement system and procedures become familiar with time, assumptions on the overall uncertainty can be made. In order to evaluate Eq. 12, the $\sigma^2_{bj}$ value of each channel $j$ should be known.

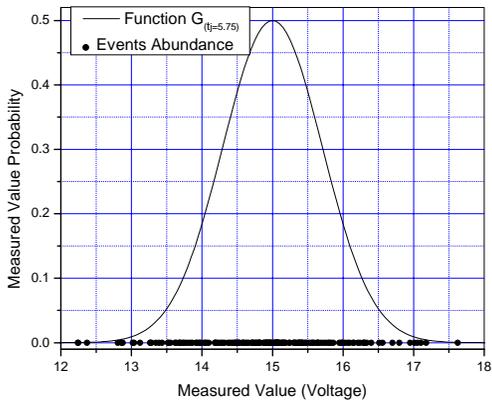

Figure 3. A sequence of data $m_j$ acquired at time $t_j$ = 5.75 s, when the same experiment is repeated in the same controlled conditions, is shown over the $x$ axis. This sequence is described by a Gaussian function [$G(t_j$=5.75 s)] with centroid in the value $\alpha f(t_j)$ = 15. Events are measured more frequently in the $x$ axis nearby of the maximum of the Gaussian function.

## 3. Application and evaluation of the developed method.

In the following we apply this methodology in order to quantify the intensity of a given known signal affected by Gaussian noise. In order to use the method, there is no need to know the place in the sequence of data where the signal should appear, since Eq. 12 is expressed as a function of a time-lag $\tau$.

We need to know, as well as possible, the sequence of data that characterize the sought function $f(t_j)$, and the variance ($\sigma^2$) of the noise, at each channel $t_j$.

In Fig. 4 we show 1) a function $f(t_j)$, 2) a sequence of Gaussian random data with $\sigma$ = 1, and 3) the composition of both sequences, when the function $f(t_j)$ is affected by a parameter $\alpha$ = 0.14. By evaluation of Eq. 12, over the sequence of the Gaussian random data, the $\Delta\alpha$ calculated is 0.045. According to the usual criterion of detection limit, the signal is effectively determined when the $\alpha$ value is at least three times $\Delta\alpha$ ($\alpha \geq 3*\Delta\alpha$; $3*0.045=0.14$). That is, the set of data represented by triangles shown in Fig.4, shows the limit of detection of the function $f(t)$ affected by a Gaussian noise with $\sigma$ = 1.

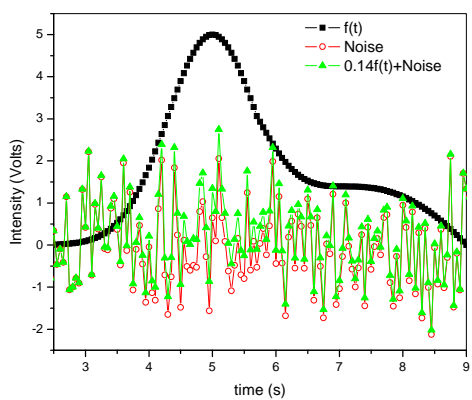

Figure 4. Sequences of data for 1) a function $f(t)$ (-■-); 2) a Gaussian background with $\sigma$ = 1 (-○-); and 3) the composition of the background plus $0.14*f(t)$ (-▲-), where the detection limit of the sought function is established.

## 4. Interpretations of the $\alpha$ parameter.

A consideration on the reading of the parameter $\alpha$ is yet to be carried out. Obtaining an *a priori* high $\alpha$ value would mean that the desired function appears with great intensity in the sequence of acquired data; whereas a small one would imply the opposite. However, this interpretation may lead to the wrong conclusions. For example Fig. 5 shows that the function $f(t)$ being quantified (square points) holds no strict relation with the $m_1$ sequence of data evaluated (circle points), nevertheless a high $\alpha$ value is obtained (the function $f(t)$ should be "expanded" by an factor of $\alpha$ = 1.96 to "fit" into the acquired data). On the other hand, another $m_2$ sequence of data (triangle points) shows a good similitude with $f(t)$, but being in most cases smaller that the function $f(t)$, produces a small $\alpha$. (the function $f(t)$ should be "compressed" by an factor of $\alpha = 0.45$ to "fit" into the acquired data)

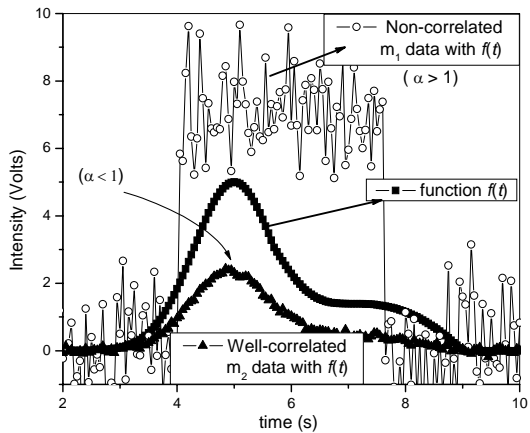

Figure 5. Sequences of data for a function $f(t)$ (-■-) and two sequences of measured data, $m_1$ and $m_2$. The sequence $m_1$ (-○-, circle points) is not well correlated with $f(t)$, but produces an $\alpha$ greater than 1. On the other hand, the sequence $m_2$ (-▲-, triangle points) is well correlated with $f(t)$ but produces an $\alpha$ smaller than 1.

To avoid these double meanings in the interpretation of the $\alpha$ parameter, we propose in first place to evaluate 1) the degree of similitude between the function and data, to later 2) determine the intensity of the sought function in the sequence of data.

If the sought function is expressed as a sequence $f=\{f_1, f_2, …, f_n\}$, and the gathered data are represented as: $m=\{m_1, m_2, …., m_n\}$, then we propose to normalize each sequence in order to quantify the degree of "similitude", by dividing each element by a factor $\zeta$ defined as:

$$\zeta_f = \sqrt{\sum_{j=1}^{n} f_j^2}$$, for the sequence defining the function,

$$\zeta_m = \sqrt{\sum_{j=1}^{n} m_j^2}$$, for the sequence defining measured data.

The new sequences obtained are expressed as:

$$f_N = \left\{ \frac{f_1}{\sqrt{\sum_{j=1}^{n} f_j^2}}, \frac{f_2}{\sqrt{\sum_{j=1}^{n} f_j^2}}, …., \frac{f_n}{\sqrt{\sum_{j=1}^{n} f_j^2}} \right\}$$ for the function, and

$$m_N = \left\{ \frac{m_1}{\sqrt{\sum_{j=1}^{n} m_j^2}}, \frac{m_2}{\sqrt{\sum_{j=1}^{n} m_j^2}}, …., \frac{m_n}{\sqrt{\sum_{j=1}^{n} m_j^2}} \right\}$$ for the sequence of data.

In the case that measured data $m$ is a proportion of the function $f$, that is, $m=xf$, (where $x\neq 0$) the "normalized" sequences are expressed as:

$$m_N = \left\{ \frac{xf_1}{\sqrt{\sum_{j=1}^{n} x^2 f_j^2}}, \frac{xf_2}{\sqrt{\sum_{j=1}^{n} x^2 f_j^2}}, …., \frac{xf_n}{\sqrt{\sum_{j=1}^{n} x^2 f_j^2}} \right\} \qquad (13)$$

which can be expressed as:

$$m_N = \left\{ \frac{f_1}{\sqrt{\sum_{j=1}^{n} f_j^2}}, \frac{f_2}{\sqrt{\sum_{j=1}^{n} f_j^2}}, ...., \frac{f_n}{\sqrt{\sum_{j=1}^{n} f_j^2}} \right\} = f_N \tag{14}$$

That is, the acquired normalized data is equal to the normalized function.

The normalized sequences for the function and for the acquired data are shown in Fig. 6. Evaluation of Eq. (12) for the new function and both data, produce $\alpha_N$ values, which will be widely discussed.

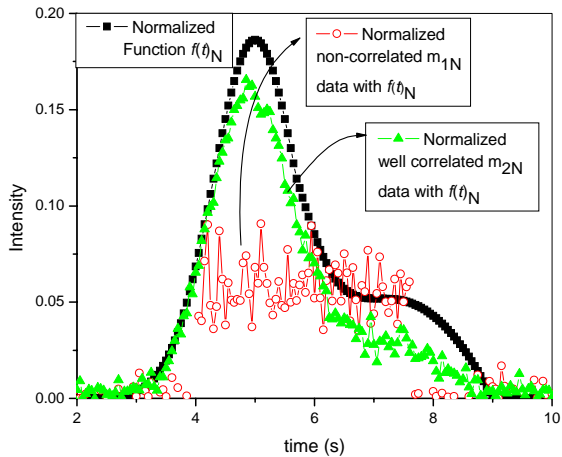

Figure 6. New sequences of normalized function $f(t)_N$ (-■-); and two sequences of normalized data, $m_{1N}$ and $m_{2N}$. The sequence $m_{1N}$ (-○-, circle points) is not well correlated with $f(t)_N$, but in this case produces an $\alpha_N$ close to zero ($\alpha_{1N}=0.27$). On the other hand, the sequence $m_{2N}$ (-▲-, triangle points) is well correlated with $f(t)_N$ which produces an $\alpha_N$ value close to 1 ($\alpha_{2N}=0.89$).

The evaluation of the parameter $\alpha_N$ when the acquired data ($m$) is a proportion of the sought function ($f=xm$), produce the values: $\alpha_N =\pm1$ (see Eqs. 13 and 14, and Fig. 7). So, the values of $\alpha_N$ can be understood as the result of a dot product (an inner product) in an Euclidean space. Since the formulation developed quantifies the degree of similitude between two signals through one single parameter, it is especially suited for finding specific structure in the data, for pattern matching problems.

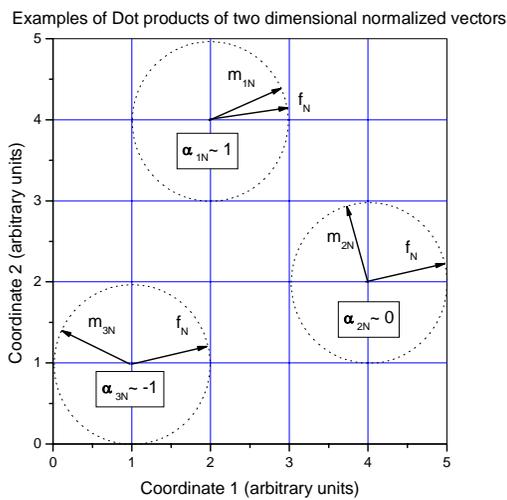

Figure 7. Three typical examples of dot products of two dimensional normalized vectors. If the vectors are approximately parallel or anti-parallel the values of $\alpha_N$ are close to $\pm1$ (data $m_{1N}$ and $m_{3N}$). If the vectors are approximately perpendicular, the value of $\alpha_N$ is close to zero (data $m_{2N}$).

## 5. Application examples.

The developed method was applied to the male voice recognition from recordings of the five Spanish vowels of a male individual, at a frequency of 11250 Hz.

*5.1 Data Acquisition.*

The five Spanish vowels from a Chilean male individual were acquired, using an XM2000S Behringer microphone and the CoolEdit Pro program, at a frequency of 11,250 Hz. Each vowel was recorded for about 2 s.

*5.2 Data Analysis.*

We have applied the developed methodology to the acquired data through an "ad hoc" Fortran 90 program. This algorithm produces the total number of $\alpha_N$ from a given function within a sequence of data. The results obtained were depicted and analyzed.

The sequences of data for each vowel were recorded for about 2 s. The procedure done in order to obtain the characteristic signal of each vowel is described in the following (all the vowels were handled in the same way, so we only will describe the data analysis procedure done over the vowel "A".):

The $m$ data for the vowel "A", $m_A$ was subdivided into five partitions, as $m_A^1,...,m_A^5$, as is depicted in Fig. 8. Each $m_A^i$ sequence contains $n = 1000$ points of data, that is, they can be represented as: $m_{A1}=\{m_A^1{}_1, m_A^1{}_2, ...., m_A^1{}_{1000}\}$.

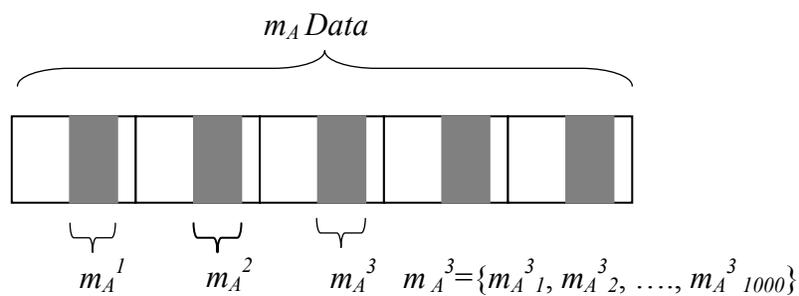

Figure 8. Five partitions of the data corresponding to the vowel "A". Each partition is composed by $n=1000$ points of data.

Each one of the five partitions $m_A^i$ is susceptible to be considered as a test function $f$, into the total sequence of data $m_A$, or in the data for the rest of the vowels ($m_E,.., m_U$). But two important questions still remain, like: ¿which is the best partition to be considered as a sought function? and, ¿can each partition be again subdivided in order to obtain the most representative information related to the vowel "A"? In order to determine the best representative partition and the best subdivision within the partition, we applied the following procedure: 1) For the first partition ($m_A^1{}_k$; $1 \leq k \leq 1000$) we determined a sequence of $n = 1000$ possible sought functions $f_n$, to be evaluated in the sequence of data $m_A$ ($m_{Aj}=\{m_{A1}, m_{A2}, ...., m_{Aj}\}$, $m_{Aj}$, $1 \leq j \leq l$, being $l$ the total number of data for the vowel "A").

For the first partition, the sequence of $n$ possible test functions $f_n$ can be described as: $f_1 = \{m_A^1{}_1\}$; $f_2 = \{m_A^1{}_1; m_A^1{}_2\}$; ....; $f_n = \{m_A^1{}_1; m_A^1{}_2; ...; m_A^1{}_n\}$. For each one of the possible sought functions $f_n$, a number of $(l-k)$ $\alpha_N$ can be computed, being $k$ the length of the function $f$. This procedure is depicted in Fig. 9.

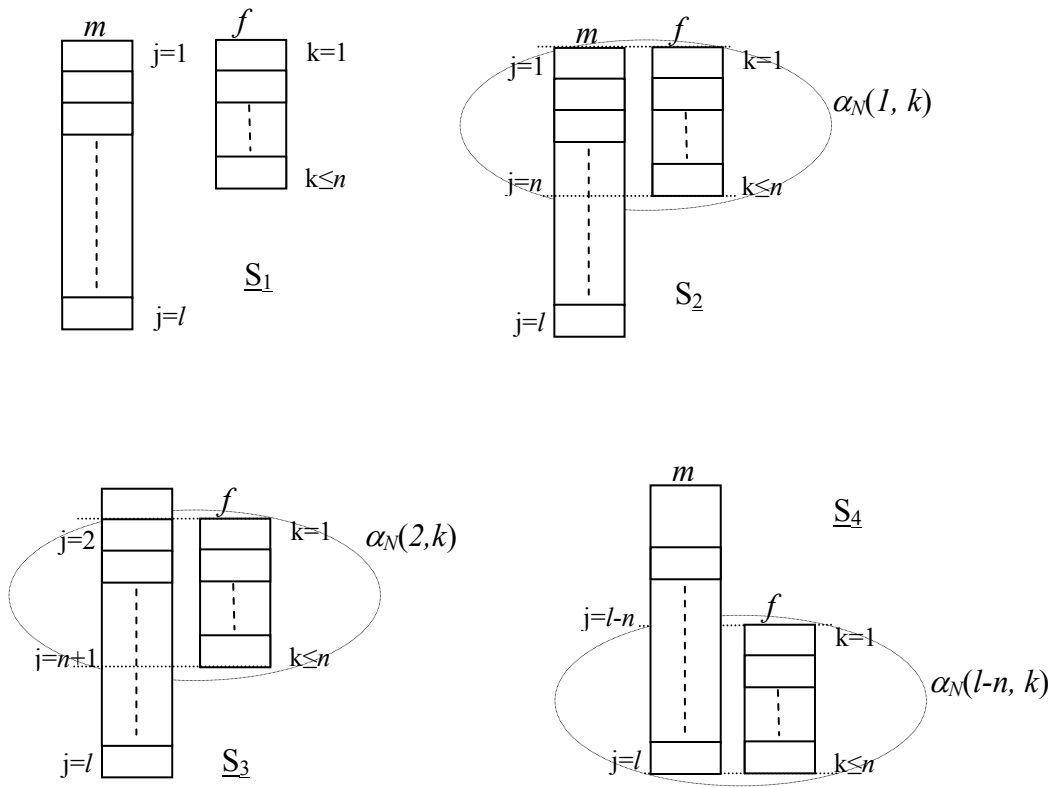

Figure 9. Scheme for the evaluation of the $\alpha_N$. If the data has a length *l* and the test function has a length *n*, a number (*l-n*) of $\alpha_N$ can be computed, as is depicted in the Stages 1, 2, 3 and 4. When this procedure is completed, the length of the function *f* is increased in one datum, and a new number of $\alpha_N$ is computed.

## 6. Results

For a function $f_k$, the number of $\alpha_N$ that can be computed within the data $m_A$ with length *l*, is *(l-k)*. Each one of the values of $\alpha_N$ indicates the degree of similitude between the function $f_k$ and a portion of the data $m_A$, with length *k*. This evaluation is shown in the Figs. 10, where the number of $\alpha_N$ greater than 0.98 (absolute values) is computed, as a fixed $f_k$ is "traveling" along the data $m_A$. Once this evaluation is concluded, the test function $f_k$ is increased in size in one datum, and the new $f_{k+1}$ is again defined. Now, a new number of $\alpha_N$ greater than 0.98 is computed along the data $m_A$, as is shown in Fig. 10.

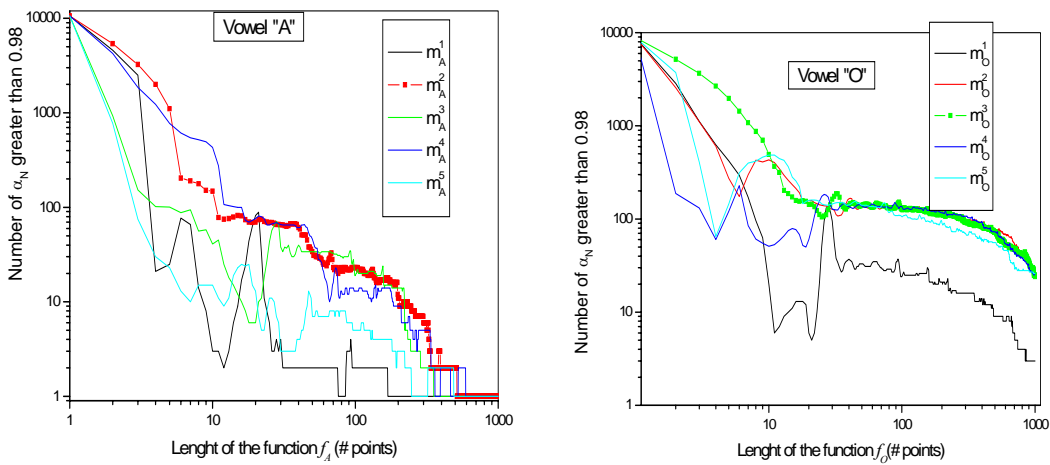

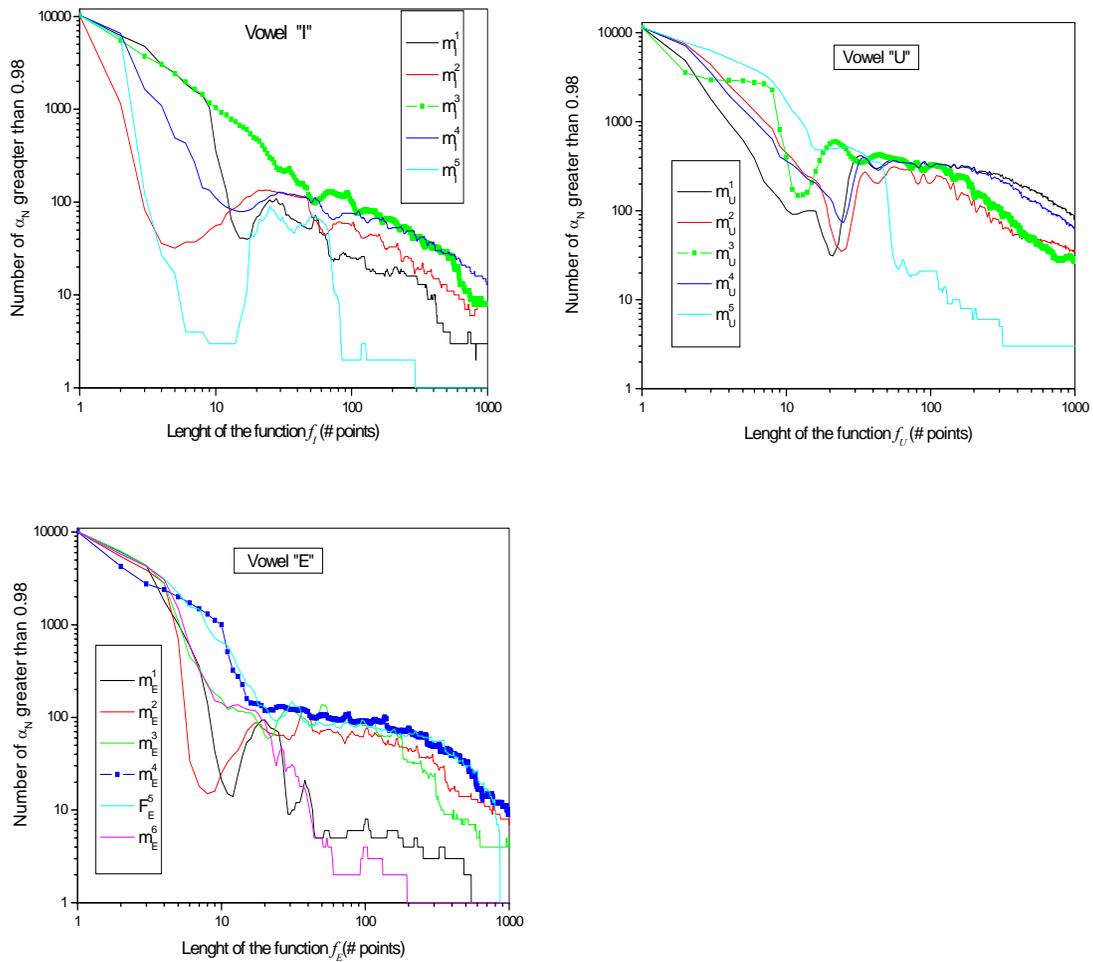

Figure 10. In each of the five graphics (one for each vowel) it is described the selection procedure of the most representative partition of data, in order to determine the best function $f$ for each vowel. The number of $\alpha_N$ greater than 0.98 (absolute values) is computed, as a fixed $f_k$ is "traveling" along the data $m_A$. Once this evaluation is concluded, the function $f_k$ is increased in size in one datum, and the new $f_{k+1}$ is again evaluated along the data $m_A$. In each graph, the data selected as the most representative for each vowel ( function $f$ ), is depicted with square points ( -■-).

Each partition $m_A^i$ represents a test function for the calculation of $\alpha_N$ values within the same signal. The $\alpha_N$ values give a criterion to select the best characteristic test function for each vowel. The most representative function would be the one that finds a greater number of $\alpha_N$, as the length of the function is increased. In all cases, intermediate partitions show a greater amount of normalized alphas through the length of the $m$ data.

Once the data selected as the most representative for each vowel is defined (square points, -■- in Fig. 10), each characteristic function was tested in itself and in the rest of the vowels. We have to determine whether the method can find a characteristic function length (function $f$), where every vowel is recognized and clearly differentiated from the other four vowels. The same comparison procedure was made in all the vowels. (See Fig. 11)

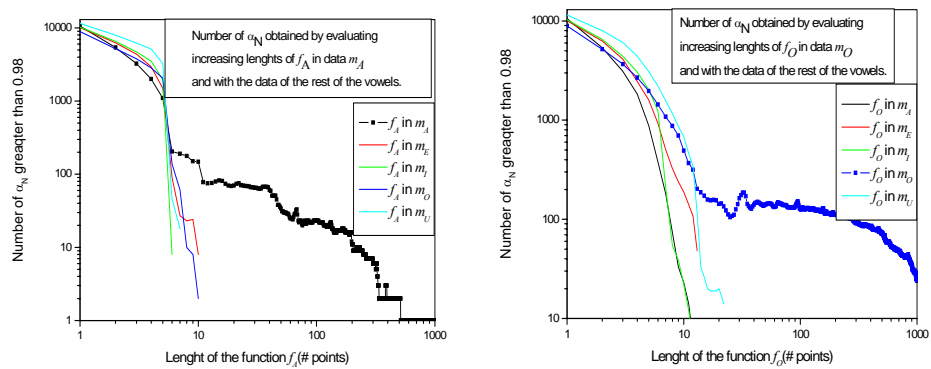

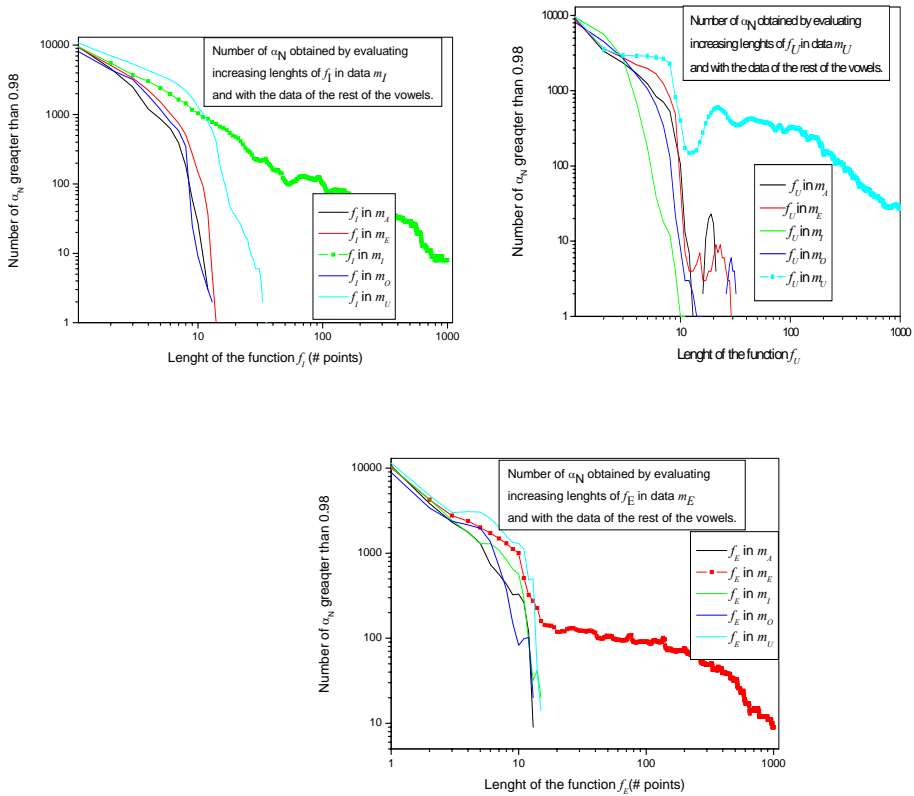

Figure 11. Number of $\alpha_N$ greater than 0.98, versus the length of the selected function $f$ for each vowel, tested against itself and with the data $m$ of the rest of the vowels.

As is observed from Fig. 11, the proposed method shows that after a certain minimum length of the characteristic function ($f$) for each vowel, it is possible to find similitude with itself and a remarkable discrepancy with the rest of the vowels (the number of $\alpha_N$ decreases dramatically) for an estimated length of 40 points (~2 $10^{-3}$ s).

## 7. Conclusions.

We have developed a methodology based on cross correlation maximization and statistical concepts, for time series analysis. This formulation opens new possibilities in the analysis of this kind of data, which were traditionally analyzed by time and frequency domain methods.

Also, we derived a new methodology which quantifies the extent of similitude between a known signal ($f(t)$) within a group of data, by means of a single parameter, $\alpha$. In order to determine a specific pattern in a given data (the $\alpha(\tau)$ value, normalized or not, with its corresponding confidence interval ($\Delta\alpha(\tau)$) the formulation proposed is evaluated analytically. The $\alpha$ values determined and their respective uncertainties are objective values of the presence of the sought function $f(t)$. Since the formulation developed quantifies the degree of similitude between two signals through one single parameter, it is especially suited for finding specific structure in the data, for pattern matching problems e.g detecting small signals in a strong background. The values of the $\alpha_N$ parameter can be understood as the result of a dot product (an inner product) in an Euclidean space.

This formulation is totally general, since the sought function $f(t)$ can be defined analytically, or by a composition of analytical functions, or simply by a sequence of data. The studied phenomenon can be non-stationary, linear, non-linear, chaotic, etc. For this formulation, data processing strategies such as filtering, noise reduction or smoothing techniques are unnecessary. Also are unimportant the evaluations of a) the complexity of the data, b) the presence or absence of fractal correlation properties or c) the magnitude of spectral power in different spectral bands. It only requires that the data be repetitive in some way in time (its intensity could not be considered), it is especially suited for pattern matching problems.

According to the studies carried out in this work, for pattern matching in time series in a group of data through a single parameter, it is possible to establish a number of conclusions related to the statistic methodology used and the obtained results in application of human voice recognition. These findings are separately presented and evaluated; however, a great extent of the conclusions may be extended and general, for later applications where analyzing similitude of a given signal within a group of data is required. From the gathered results and the studied sample data it is possible to state that:

According to the sample data, recording of the 5 vowels produced by a male voice at a 11.250 Hz frequency during an estimated 2 s, the same decreasing tendency is observed (where a characteristic form is implicit) of $α$ number as the length of the function for each partition in the same vowel is increased.

By selecting a characteristic function for each vowel, and by comparing it to itself and to the other vowels, graphics show that for different vowels the number of $α$ decreases dramatically, whereas for the characteristic function there is still a considerable amount of $α_N$ values.

Even though a goodness of test was not carried out over the data analyzed, the proposed method shows that after a certain characteristic length of the test function for each vowel, it is possible to find similitude with itself and a remarkable discrepancy with the rest of the vowels for an estimated length of 30 points (~2 $10^{-3}$ s).